\documentclass[conference]{IEEEtran}
\IEEEoverridecommandlockouts
\usepackage{cite}
\usepackage{amsmath,amssymb,amsfonts}
\usepackage{algorithmic}
\usepackage{graphicx}
\usepackage{textcomp}
\usepackage{xcolor}
\usepackage{bm}
\usepackage{comment}
\usepackage{xspace}
\def\BibTeX{{\rm B\kern-.05em{\sc i\kern-.025em b}\kern-.08em
    T\kern-.1667em\lower.7ex\hbox{E}\kern-.125emX}}

\newcommand*{\name}{ET$^3$\xspace}
\newcommand*{\q}[1]{``#1''}
\begin{document}

\title{Visual Anomaly Detection in Event Sequence Data\\}

\author{\IEEEauthorblockN{Shunan Guo}
\IEEEauthorblockA{Department of Software Engineering\\
East China Normal University\\
Email: g.shunan@gmail.com}
\and
\IEEEauthorblockN{Zhuochen Jin}
\IEEEauthorblockA{College of Design and Innovation\\
Tongji University\\
Email: chjzcjames@gmail.com}
\and
\IEEEauthorblockN{Qing Chen}
\IEEEauthorblockA{College of Design and Innovation\\
Tongji University\\
Email: jane.qing.chen@gmail.com}
\and
\IEEEauthorblockN{David Gotz}
\IEEEauthorblockA{School of Information and Library Science\\
University of North Carolina at Chapel Hill\\
Email: gotz@unc.edu}
\and
\IEEEauthorblockN{Hongyuan Zha}
\IEEEauthorblockA{Department of Software Engineering\\
East China Normal University\\
Email: zha@cc.gatech.edu}
\and
\IEEEauthorblockN{Nan Cao}
\IEEEauthorblockA{College of Design and Innovation\\
Tongji University\\
Email: nan.cao@gmail.com}
}

\maketitle

\begin{abstract}
Anomaly detection is a common analytical task that aims to identify rare cases that differ from the typical cases that make up the majority of a dataset. When applied to the analysis of event sequence data, the task of anomaly detection can be complex because the sequential and temporal nature of such data results in diverse definitions and flexible forms of anomalies. This, in turn, increases the difficulty in interpreting detected anomalies. In this paper, we propose an unsupervised anomaly detection algorithm based on Variational AutoEncoders (VAE) to estimate underlying normal progressions for each given sequence represented as occurrence probabilities of events along the sequence progression. Events in violation of their occurrence probability are identified as abnormal. We also introduce a visualization system, EventThread3~(\name), to support interactive exploration and interpretations of anomalies within the context of normal sequence progressions in the dataset through comprehensive one-to-many sequence comparison. Finally, we quantitatively evaluate the performance of our anomaly detection algorithm and demonstrate the effectiveness of our system through a case study.
\end{abstract}


\vspace{-0.1cm}
\section{Introduction}
\label{sec:intro}
\maketitle
Anomaly detection is a common task for event sequence data analysis as it often contributes to the discovery of critical and actionable information~\cite{chandola2009anomaly}. Effective use of event sequence data can require identifying sequences that deviate from the typically occurring behavior~\cite{patcha2007overview}. For example, a doctor may be interested in finding patients whose postoperative response is different from other patients who have had the same surgery, so that the doctors can provide personalized care plans for similar patients in the future. A variety of techniques, including traditional statistical models~\cite{qayyum2005taxonomy, xiong2012energy}, supervised or semi-supervised approaches~\cite{liu2008isolation}, and unsupervised methods~\cite{munz2007traffic} have been applied to detect anomalies in event sequences. However, due to the temporal characteristics of event sequence data and the black-box nature of machine learning models, it is challenging to interpret anomalous sequences once identified. For analysts to derive actionable insights, they must be able to understand how anomalies are different from ``normal'' sequences, which event or series of events characterize the anomaly, and which events suggest actions that could help avoid such cases in the future.

In this paper, we propose an unsupervised anomaly detection model for event sequence data that builds upon LSTM-based Variational AutoEncoders (VAE). VAE use a probabilistic encoder for modeling the distribution of the latent variables. Such probabilities give more principled criteria for identifying anomalies and do not require model-specific thresholds, which in turn, better facilitate objective judgments for deciding the boundary of anomalous sequences compared to other unsupervised algorithms. We train the model to learn a latent representation for each event sequence and identify anomalous sequences based on their deviation from the overall distribution. A mean sequence is computed from the reconstruction probabilities for each sequence detected as an anomaly, which shows the occurrence probabilities of events in normal circumstances, representing a corresponding ``normal'' sequence progression for the anomaly. For example, a patient having internal bleeding should normally be sent to emergency for surgery, thus the reconstruction probabilities shall identify surgical events with high probabilities after hospital admission. To facilitate interpretation of the anomaly, we also present an interactive graphical interface system to compare the anomaly sequence with a collection of normal sequences through one-to-many comparison mechanism and uncover their critical differences through a specifically designed comparison glyph.
\section{Related Work}
\label{sec:related}
Anomaly detection has been extensively studied over the past years\cite{chandola2009anomaly}. Methods for anomaly detection can be broadly categorized into tensor-based algorithms\cite{chen2015uncertainty}, statistics-based algorithms\cite{rousseeuw2005robust}, classification-based algorithms\cite{liu2008isolation}, and neighbor-based or distance-based algorithms\cite{breunig2000lof}. More recent work with deep learning-based anomaly detection (DAD) algorithms has been developed to pursue better performance. Types of DAD models include unsupervised (e.g., autoencoder, generative adversarial, variational), semi-supervised (e.g., reinforcement learning), hybrid (e.g., feature extractor + traditional algorithms)\cite{erfani2016high}, and one-class neural networks\cite{chalapathy2018anomaly}.  In this work, we leverage VAE which can both deal with large volumes of unlabeled data, and identify anomalous patterns with probability measures\cite{an2015variational}. Furthermore, we utilize the reconstruction probabilities to generate a output close to the original input sequence and provides latent feature vectors for each event sequence in the dataset.

Incorporating human domain knowledge through interaction can benefit the anomaly detection process, especially when the boundary between normality and abnormality is not precisely defined. To this end, researchers have developed many visual anomaly detection tools \cite{cao2018voila,thom2012spatiotemporal}. This includes methods for the detection of anomalous user behaviors from sequence data\cite{bock2015visual}. Chae et al.\cite{chae2012spatiotemporal} applied traditional control chart methods together with seasonal trend decomposition to extract outliers. Thom et al.\cite{thom2012spatiotemporal} introduced a visual analysis system to monitor for anomalous bursts of keywords. More recently, FluxFlow\cite{zhao2014fluxflow} was developed to reveal and analyze anomalous information processes in social media. Although these systems are often designed to help detect anomalous points, few approaches focus on identifying anomalous sequences or on the comparison between the detected outliers and ``normal'' sequences. To enhance the interpretability of the analyzed results, in \name, we provide an interactive one-to-many comparison between the anomalous sequence and normal progressions.
\section{VAE-Based Anomaly Detection}
\label{sec:analysis}

\subsection{LSTM-Based Variational AutoEncoder}

We first introduce the structure of the Sequence-to-Sequence VAE model. The model contains two modules: the VAE encoder and the VAE decoder. Both modules are designed using Recurrent Neural Networks to better extract sequential patterns from event sequence data. In particular, the encoder captures the latent distribution of sequences and the decoder inversely restores the distribution to estimate the occurrence probabilities of events in each time slot.

\textbf{VAE Encoder.} The encoder is trained to abstract the input sequence $ \{X = \bm{x}_i\}_{i=1}^{n}$ into a low-dimensional latent feature vector that describes a sequential distribution of events occurring in the sequence(as shown in Fig.~\ref{fig:model_anomaly}(1)). In this input, $n$ is the length of the sequence and $\bm{x}_{i} \in \{0,1\}^{|E|}$ is the multi-hot encoding of the events in event set $E$ occurring in the $i$-th time step. Each coordinate represents an event type, which is marked 1 if the corresponding event occurs in the $i$-th time step, or 0 otherwise. After feeding the multi-hot vectors into the corresponding layer of RNN, the state of the entire sequence is extracted and represented in the hidden state vector $\bm{h}_{enc}$ of the last layer, which is denoted as $\bm{h}_{enc} = encoder(X)$.
The hidden state vector $\bm{h}_{enc}$ is projected into vector $\bm{\mu}$ and $\bm{\delta}$ to parameterize a normal distribution, representing the mean value and standard deviation of the normal distribution respectively. To take the variability of the latent space into account (i.e., to represent the diversity present in normal cases), we draw a low-dimensional latent vector $\bm{z}$ by randomly sampling from the distribution and use this vector as a representative of the original distribution for subsequent decoding.

\textbf{VAE Decoder.} In the decoder, we reconstruct the input sequence from the extracted latent feature vector $\bm{z}$. Specifically, $\bm{z}$ is fed to each layer of the RNN to estimate the probability distribution of events for each time slot. We formally define the decoding procedure as $ X^{'} = decoder(z) \label{eq:decoder}$, where $X^{'}$ is a sequence of probability distributions denoted as $X^{'}= \{\bm{x}_{i}^{'}\}_{i=1}^{n}$, and the element $x_{i,j}^{'}$ in $\bm{x}_{i}^{'} \in R^{|E|}$ represents the occurrence probability of j-th event at the i-th time step.

\textbf{Training Process.} 
We train the model with a goal of narrowing the gap between the original input sequence and its reconstruction, which can be formally defined as minimizing the following loss function: 
\vspace{-0.5em}
\begin{align}
L &= L_{r} + w_{kl} \cdot L_{kl}  \\
L_{r} &= \frac{\sum\limits_{i=1}^{n} \sum\limits_{j=1}^{|E|} (w_{e_{j}}x_{ij}log(x_{ij}^{'}) + (1-x_{ij})log(1-x_{ij}^{'}))}{-n}\\
L_{kl} &= - \frac{1}{M_{z}}\sum_{i=1}^{M_{z}}(1+log(\sigma_{i}^{2})-\mu_{i}^{2}-\sigma_{i}^{2})
\end{align}
\vspace{-0.4em}

The first term $L_{r}$ is the reconstruction loss which calculates the weighted cross entropy between $x_{i,j}$ and $x_{i,j}^{'}$, indicating an event-level difference between the reconstruction and the original input with respect to the $j$-th event at the $i$-th time step. In particular, a parameter $w_{e_{j}}=1/log(n_{j})$ is introduced to reduce the marginal importance of high-frequency events so as to address the issue of skewed dataset, where $n_j$ is the number of occurrences for event $e_j$.
The second term $L_{kl}$ is the Kullback-Leibler Divergence Loss which estimates a distribution-level difference between the distribution of the latent vector $\bm{z}$ and a normal distribution $N(0,1)$, where $M_z$ is the dimension of the latent vector $\bm{z}$. These two terms are balanced with a parameter $w_{kl}$.

\textbf{Parameter Settings.} 
Both the encoder and decoder employ LSTM units~\cite{hochreiter1997long} with 300 hidden nodes. We set the dimension of the latent vector to 16. The parameter $w_{kl}$ adaptively increases from 0.1 to 0.5 during the training process to make sure the reconstruction loss is optimized with high priority. Moreover, we optimize the loss function with the Adam optimizer~\cite{kingma2014adam} with training data batch size of 80 for each training step.  We train the model on an Nvidia Tesla K80 graphics card. Each training epoch takes approximately 10.5 seconds on average.

\textbf{Anomalous Sequence Detection.}
After training the model, we employ the latent vector $\bm{z}$ of each input sequence to detect anomalous sequences in the dataset to calculate the degree of anomaly for each sequence in the latent space using the Local Outlier Factor (LOF)~\cite{he2003discovering}(as shown in Fig.~\ref{fig:model_anomaly}(2)). Normal sequences should group within a dense space with smaller LOF scores, while instances in sparse areas will have larger LOF scores and will be identified as outliers.

\subsection{Anomalous Event Analysis}
\label{sec:event_analysis}
\begin{figure}
    \centering
    \includegraphics[width=0.9\linewidth]{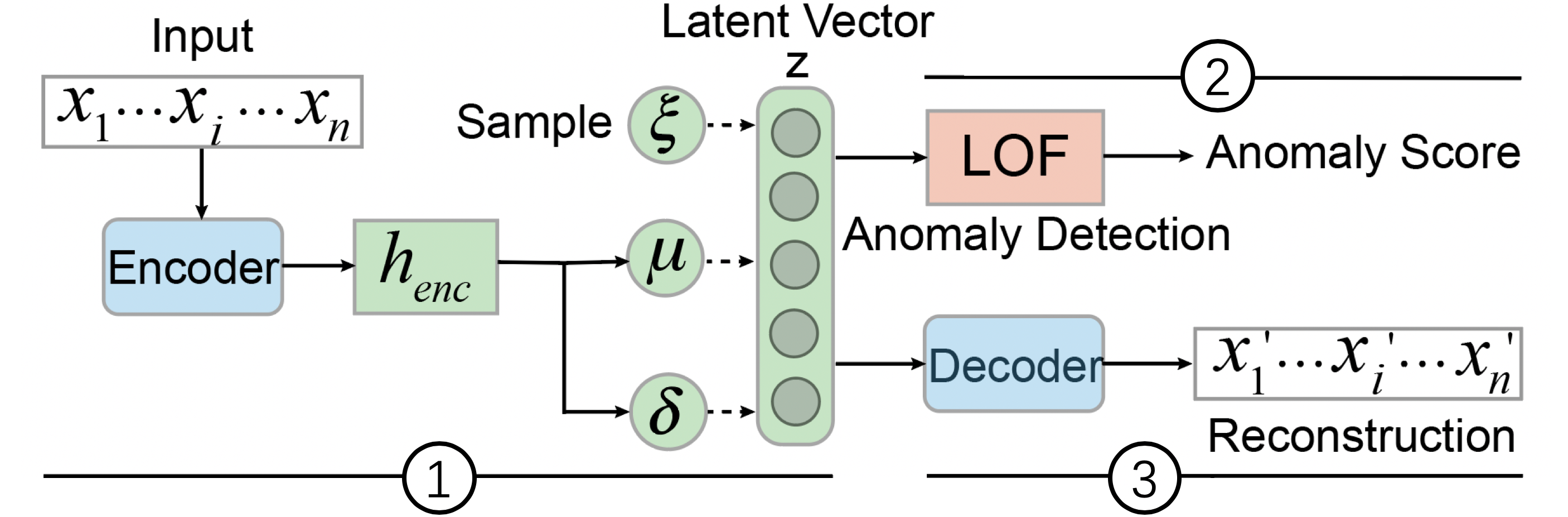}
    \vspace{-1em}
    \caption{Schematic diagrams of the model, (1) the VAE model to obtain the latent vector of the input sequence, (2) anomaly detection of the overall sequence, and (3) anomalous event detection based on the reconstruction of the input sequence.}
    \label{fig:model_anomaly}
    \vspace{-1.5em}
\end{figure}

To facilitate the interpretation of sequence anomalies, we further identify anomalous events that contribute to sequence abnormality by analyzing the reconstruction probabilities(as shown in Fig.~\ref{fig:model_anomaly}(3)). As we assumed that the majority of the sequences are normal, the reconstruction probabilities shall be similar to the normal progression of sequences, and the training objective ensures that the reconstruction probabilities are also similar to the original input sequence. Thus, the reconstruction probabilities of the anomalous sequences can be used to infer a \textit{mean sequence} that represents an expected ``normal'' progression for the anomalous sequence. From this, we can identify the anomaly events within the anomalous sequence that deviate from the expected normal progression.

We categorize the anomalous events into missing events (noted as $x^{mis}$) and redundant events (noted as $x^{red}$), which represent the cases where events show high occurrence probabilities in the reconstruction but do not appear in the sequence, and events that exist in the anomalous sequence but are not expected to occur, respectively. Based on this intuition, we calculate the anomaly scores for missing events and redundant events with $Pr(X=x^{mis})$ and $1-Pr((X=x^{red}))$, respectively, where $Pr(X=x)$ indicates the occurrence probability of the corresponding event derived from the reconstruction.
Consequently, events with an anomaly level higher than a user-defined threshold are identified as anomalous. The threshold is by default set as 0.6, which can be adjusted by users during an analysis via the visualization module.

\section{Visualization}
\label{sec:visual}

\begin{figure*}[t!]
\centering
\includegraphics[width=2\columnwidth]{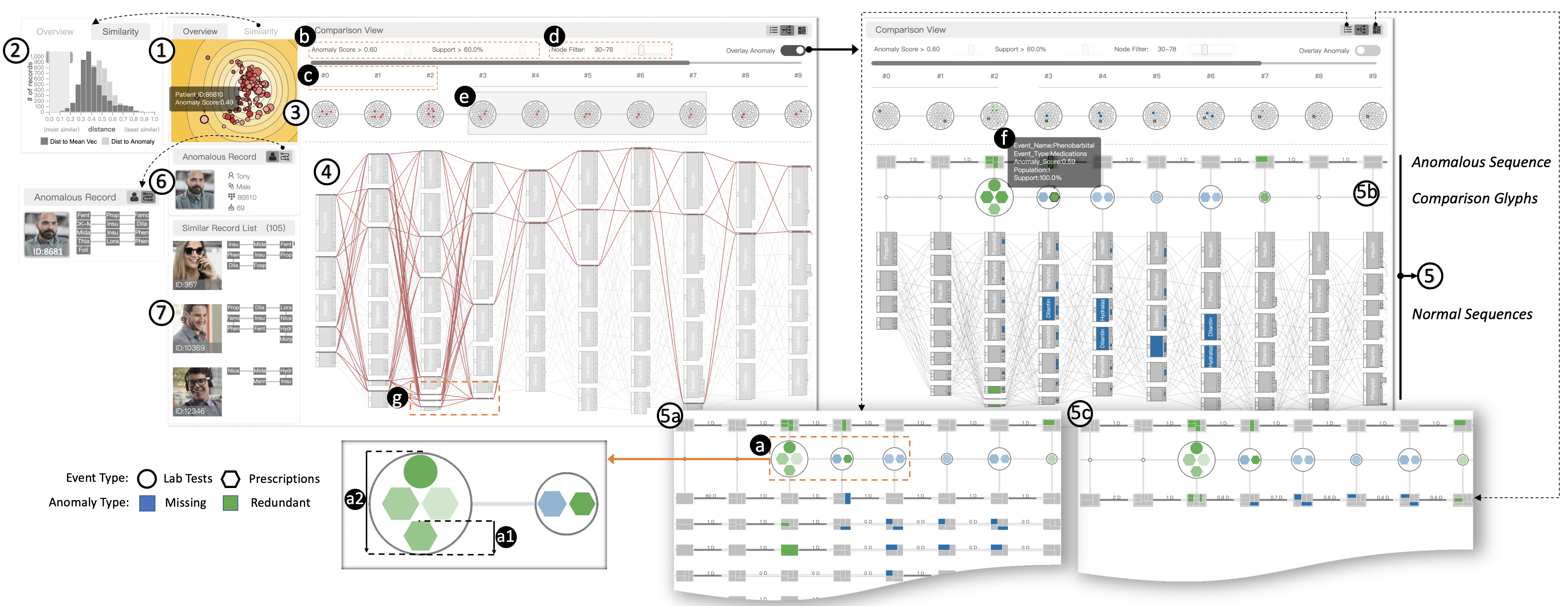}
\vspace{-1em}
\caption{The user interface of \name consists of seven key views with a comparison glyph designed to support comparison-based visual anomaly detection.}
\label{fig:interface}
\vspace{-1.5em}
\end{figure*}

\subsection{Design Tasks}
We formulated a set of design tasks to solve the key challenges in visually analyzing anomalies in event sequence.

\begin{itemize}
    \item[\textbf{T1}]\textbf{Provide an overview of the analysis scope.} To help users find anomalous sequences of interest within a collection of anomalous sequences, the system should provide an overview of all sequences detected as anomalies and illustrate their level of abnormality.
    
    \item[\textbf{T2}]\textbf{Emphasize anomalous events within the sequence.} To help quickly explore complex event sequences and uncover the reason behind an abnormality, the visualization should be designed to highlight key events that are suspicious of being anomalous. 

    \item[\textbf{T3}]\textbf{Facilitate result interpretation in context.} 
    The designed visualization should help users effectively analyze the detected anomalies within the context of the entire training set, to uncover the difference between abnormal and normal sequence progressions and facilitate reasoning about the analyzed result.
    
    \item[\textbf{T4}]\textbf{Support sequence exploration at multiple levels of granularity.} 
    Applying different levels of aggregation for a group of sequences can result in distinct interpretations of the result. To support more accurate findings, the system should support the exploration of normal sequences at different levels of granularity.
\end{itemize}

Guided by the tasks above, \name incorporates seven key views to visually analyze the anomalous sequences (Fig.~\ref{fig:interface}), which includes (1) an \textit{anomaly overview} providing an overview of all detected anomalies from which users can choose for subsequent analysis \textbf{(T1)}; (2) a \textit{similarity view}, showing the distribution of normal sequences as regard to their similarities and the selected anomaly \textbf{(T1)}; (3) a \textit{reconstruction view} presenting the occurrence probabilities of the events in each time slot~\textbf{(T3)}; (4) a \textit{flow overview} that aggregates the flow of normal sequences with the evolution of the selected anomaly overlaid at the top to show differences~\textbf{(T3)}; (5) a comparison view with three variants: (5a) \textit{sequence comparison view}, (5b) \textit{flow comparison view}, and (5c)  \textit{summarization view} that separates the flow of the anomalous sequence from the normal sequences with \textit{comparison glyphs}~(Fig.~\ref{fig:interface}(a)) emphazing the event difference~\textbf{(T2)} and displays the summarization of normal sequences in multi-level granularity~\textbf{(T4)}; (6) an \textit{anomalous record view} and (7) a \textit{similar record list} displaying low-level details of the selected anomaly and normal sequences, respectively. The system is incorporated with rich interactions to support exploratory analysis.

\subsection{Interactive Anomalous Event Analysis}
Our system is designed to facilitate interpretation of the selected anomaly in the context of the progression of normal sequences~\textbf{(T3)} via interactive one-to-many visual comparison. The comparison view is vertically divided into three regions: an \textit{anomalous sequence} at the top, a group of \textit{comparison glyphs} in the middle, and a summarization of normal sequences at the bottom. 

\subsubsection{Anomalous Sequence}
The selected anomalous sequence is displayed using a line of rectangular nodes ordered by time of occurrence. To deal with the issue of event co-occurrence and avoid event overlap, we display the sequence with a visual technique introduced in~\cite{jin2018carepre}. Specifically, concurrent events are grouped into treemaps at each time slot, and all event nodes are color-coded according to the type of anomaly. Event nodes are spaced with equal distance and connected with duration bars to reveal the span of time. The time span between events is proportional to the duration bar.

\subsubsection{One-to-Many Sequence Comparison}
\label{sec:comparison}
Our system incorporates a one-to-many sequence comparison mechanism, which allows users to validate the anomalies detected by the model by comparing the anomalous sequence with a collection of similar sequences from the normal group. This aims to help users establish confidence in the analysis result based on evidence in the dataset. The comparative analysis consists of two steps: sequence alignment and support rate calculation. In the first step, we employ a sequence alignment technique introduced in~\cite{guo2019visual} to semantically map each normal sequence to the focal anomaly based on Dynamic Time Warping (DTW)~\cite{muller2007dynamic} to address the issues of variable sequence length and progression rate for a more precise comparison of events. After sequence alignment, we compare events occurring in each time slot to calculate a \textit{support rate} for each anomalous event identified in Sec.~\ref{sec:event_analysis}. 
Intuitively, the support rate represents the proportion of normal sequences that ``support'' the corresponding event to be abnormal. More specifically, the support rate of a missing event $x^{mis}$ is the proportions of sequences that include $x^{mis}$ in the corresponding time slot, while the support rate for $x^{red}$ is the contrary.

\subsubsection{Comparison Glyph}
\label{sec:glyph}
To facilitate visual comparison, we design a \textit{comparison glyph} (Fig.~\ref{fig:interface}(a)) that highlights the anomalous events in each time slot. We encode four critical variables to help quickly identify problematic time slots and events that need further inspection: the overall abnormality of the time slot, the abnormality of each event, the type of anomaly, and the support rate for each anomalous event. Specifically, each circle inside the glyph represents an anomalous event. The size of each internal circle(Fig.~\ref{fig:interface}(a1)) indicates the anomaly score of the corresponding event derived from Sec.~\ref{sec:event_analysis}, and the size of outer circle (Fig.~\ref{fig:interface}(a2)) represents the overall abnormality at the corresponding time slot. The type of abnormality (e.g., missing event or redundant event) is distinguished with different colors, consistent with other views. The support rate of each anomalous event (Sec.~\ref{sec:comparison}) is encoded with color saturation. 

\textbf{Updating with user feedback.} To leverage analyst domain knowledge, the system allows users to interactively tweak the anomalous events displayed in the comparison glyphs.
As shown in Fig.~\ref{fig:interface}(b), users can tune the thresholds for the anomaly score and support rate that determine the conditions at which an event is identified as anomalous. Moreover, when users select a subgroup of normal sequences during the analysis, the comparison glyphs will also be updated simultaneously to reflect the support rate within the subgroup. 

\subsubsection{Multi-granular Sequence Aggregation}
To support more comprehensive one-to-many sequence comparisons, the design provides three coordinated comparison views (Fig.~\ref{fig:interface}(5a-c)). The views support comparison at different levels of aggregation \textbf{(T4)}, and transitions allow users to move smoothly from one view to another.

The \textit{sequence comparison view} displays the sequences of normal records individually, which aims to support sequence-to-sequence level comparison and efficient access to the raw data. As shown in Fig.~\ref{fig:interface}(5a), the normal sequences are displayed in a scrollable list with a consistent encoding schema as the anomalous sequence, ranked from top to bottom according to the degree of similarity. Users can select any individual sequence to update the comparison glyphs with their differences during the analysis.

The \textit{flow comparison view} (Fig.~\ref{fig:interface}(5b)) provides a progression-level summarization on all normal sequences by aggregating them into a flow-based visualization. This view aims to incorporate confidence of abnormality for anomalous events by comparing the anomalous sequence with subgroups of sequences having particular progression patterns. Specifically, identical events in each time slot are grouped into nodes, and the transition paths among events in adjacent time slots are merged into links. 
The height of each node represents the population (weighted by event co-occurrence) having the event at the corresponding time slot, with the exact number displayed in a label to the left side of each node.  Event nodes are connected with links to represent a sequence path from one event to another. Each link is consist of a duration bar and a connection line. The height of the duration bar shows the proportion of the population corresponding to the link, while the width indicates the average time gap between events. Users can select any node or link to highlight the progression pattern and narrow the comparison of a specific subgroup. 

In the \textit{summarization view} (Fig.~\ref{fig:interface}(5c)), nodes in each time slot are further aggregated into a more compact form, illustrating the highest-level summarization of the distribution of events. This view aims to support a comparison of the anomalous sequence against the overall progression of the entire set of similar records. We encode the summarized sequences in a way similar to the anomalous sequence, with the only difference that the size of each inner rectangle represents the size of the population. To allow for the analysis and exploration on a higher-level summarization of progression stages, we also leverage a recently proposed progression analysis technique~\cite{guo2019visual} to segment the anomalous sequence into different stages. Stages are marked with line segments under the identifier of the time slots (Fig.~\ref{fig:interface}(c)). Users can click on a stage identifier to merge or expand all visual elements in the main panel that align to the corresponding time slots. 

\subsection{Other Views}
The system also includes several contextual views to display auxiliary information and provide access to raw data. The \textit{anomaly overview} (Fig.~\ref{fig:interface}(1)) shows the multidimensional scaling (MDS) projection of the latent vector $z$ on a colored contour map. Each anomalous sequence is represented as a circle with the size indicating the LOF score, and the color saturation indicating the sequence length, so as to help analysts choose sequences of high anomaly degree for subsequent analysis. The \textit{similarity distribution view} (Fig.~\ref{fig:interface}(2)) displays the distribution of all normal sequences in the dataset based on their similarity to the selected anomaly, which aims to help users select a proper group of normal sequences for comparative analysis. The reconstruction probabilities (given by Equation~\ref{eq:decoder}) of the selected anomaly are shown in the \textit{reconstruction view} (Fig.~\ref{fig:interface}(3)) with the intent of providing an overview of the occurrence probabilities of events for each time slot. The reconstruction probabilities are shown as a line of circle packings arranged in time order, with the size of each circle shows the value of probability, and the color indicates different anomaly types (consistent with other views). The \textit{anomalous record view} (Fig.~\ref{fig:interface}(6)) and the \textit{similar record list} (Fig.~\ref{fig:interface}(7)) provide access to raw event sequence data of the anomalous sequence and similar sequences. These low-level details provide detailed evidence to support interpretation.

\section{Evaluation}
\label{sec:eval}
We demonstrate the effectiveness of \name's analytical model and the usefulness of visualization system through a quantitative evaluation and a case study.

\begin{figure}[t!]
\centering
\includegraphics[width=\columnwidth]{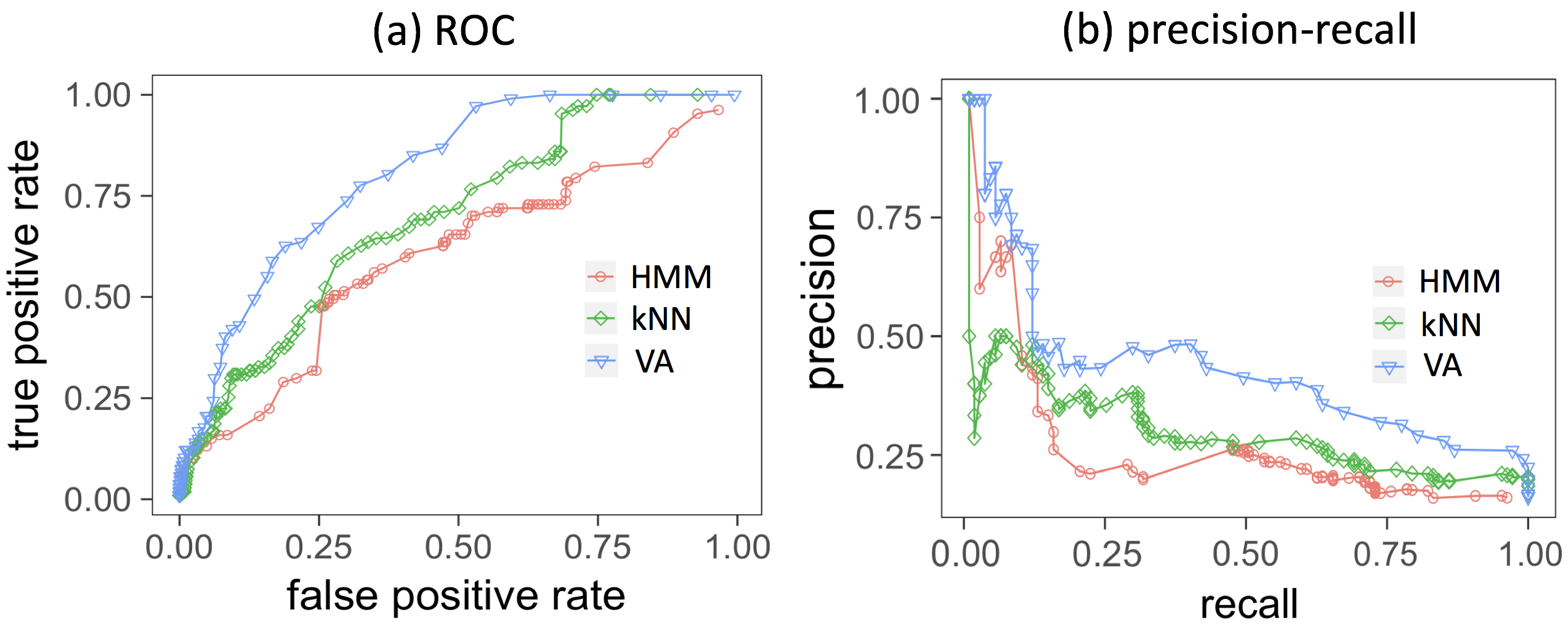}
\vspace{-2em}
\caption{Performance evaluation results of our VAE-based algorithm (VA) in comparison with two baseline methods (kNN, HMM).}
\label{fig:performance}
\vspace{-1.5em}
\end{figure}

\vspace{-0.1cm}
\subsection{Quantitative Evaluation}
We compare the performance of our VAE-based anomaly detection algorithm (denoted as VA) with two baseline methods using an intrusion detection dataset, snd-cert~\cite{forrest1996sense}. The dataset consists of sequences of operating system calls that are labeled in terms of the system state (i.e., normal or hacked) when running these operations. 

We select two representative baseline methods under the categories of kernel-based and Markovian anomaly detection techniques: Nearest Neighbor (kNN)~\cite{liao2002use} and Hidden Markov Model (HMM)~\cite{xie2009large}. Both methods have been shown efficient for detecting anomalies in event sequence data in previous research~\cite{budalakoti2006anomaly, qiao2002anomaly, zhang2003new}. 
More specifically, the longest common subsequence (LCS) was used as the distance metric in kNN and standard information retrieval metrics (precision, recall, and ROC) to evaluate model performance. 

\textbf{Evaluation Results.}
Our algorithm outperforms the baseline methods as shown in Fig~\ref{fig:performance}. 
The ROC plot (Fig~\ref{fig:performance}(a)) illustrates that VA achieves higher true positive rates when the false positive rates remain low (below 0.25) compared to the other two baseline methods. The precision-recall plot (Fig~\ref{fig:performance}(b)) shows that VA had overall higher precision than the baseline methods. The results indicate that our approach can produce a higher quality set of suspicious sequences when compared to the baseline algorithms. Using the designed visualization, the system can further support the interpretation of detected anomalies.

\vspace{-0.1cm}
\subsection{Case Study}
We applied \name to MIMIC~\cite{johnson2016mimic}, a publicly accessible critical care database with de-identified electronic health records for 46,520 patients with 12,487 event types in total. Due to the diversity of sequence progression for patients with different diseases, training with the entire database could introduce noise and produce inaccurate anomaly results. With this consideration, we selected a subgroup of 10,183 patients who were diagnosed with cardiovascular diseases to produce a more homogeneous set of sequence progressions for training. Four cardiologists (\textbf{E1--E4}, 5--8 years of domain experience each) were invited to participate in our study. Prior to the study, we asked the doctors about expected patterns of anomaly and they expressed interests in exploring anomalous medical usage within the follow-up lab test results, based on which we extracted 87 types of \textit{prescriptions} and \textit{lab events}, which were further distinguished using different shape in the visualization. 

\begin{figure}[t!]
\centering
\includegraphics[width=\columnwidth]{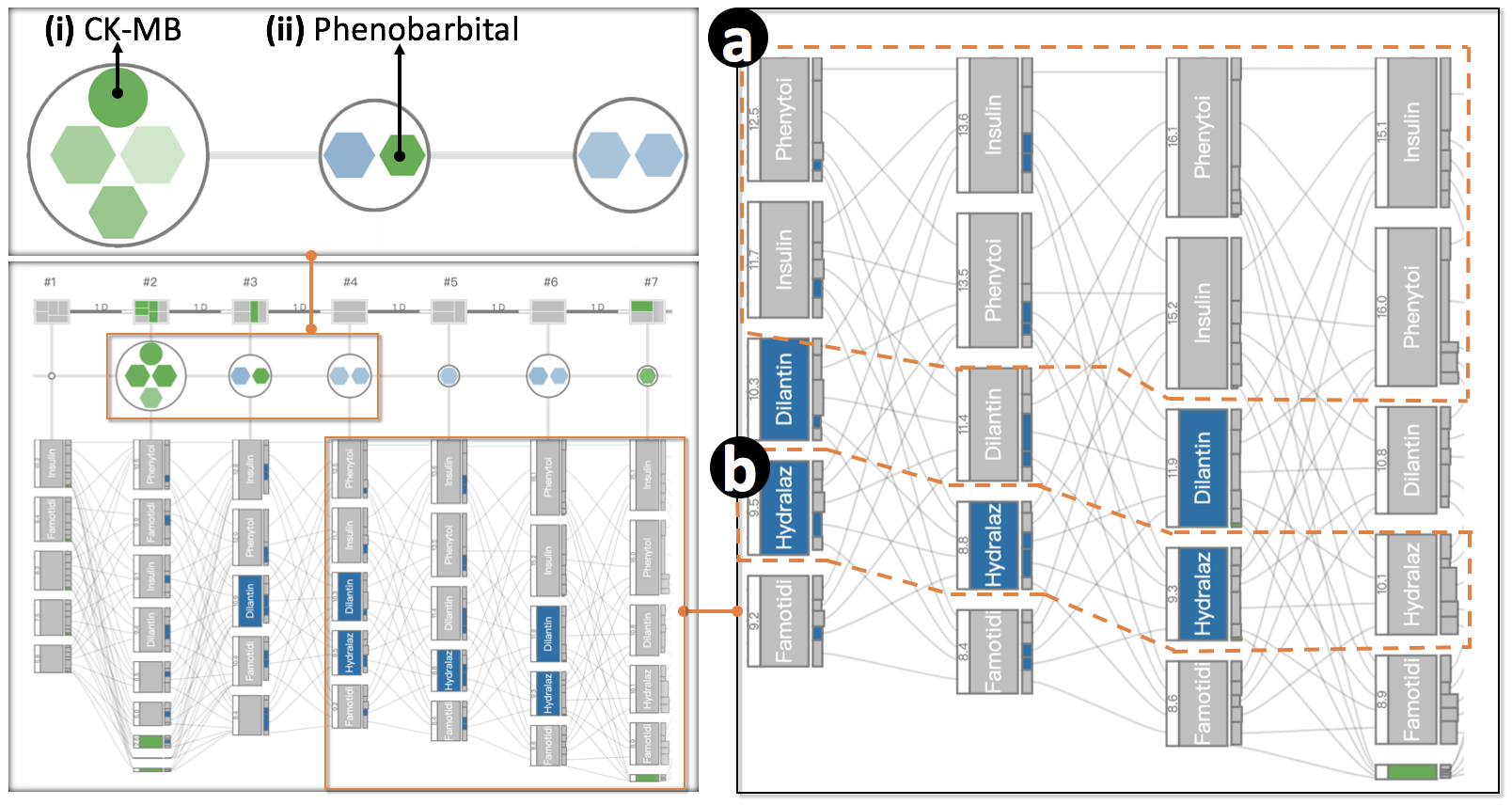}
\vspace{-2em}
\caption{The anomaly detection result of MIMIC dataset. The system identified major progression paths (a-b), from which the sequence anomaly deviates in (i) an anomalous lab test result and the (ii) misuse of a prescription drug. }
\label{fig:mimic}
\vspace{-1.5em}
\end{figure}

404 anomalous sequences were detected for subsequent analysis. The experts chose a patient who was far away from the main cluster with a relatively high anomaly score from the overview(Fig.~\ref{fig:interface}(1)), and then retrieved 105 similar patients with the distance to the mean sequence under 0.2 for subsequent analysis (Fig.~\ref{fig:interface}(2)). As they quickly scrolled the \textit{flow overview} back and forth to get a big picture of the major sequence progression paths, they found that the treatment plans for the patients were very similar with regular use of Phenytoin and Insulin (Fig.~\ref{fig:mimic}(a)), 
and speculated that all of these patients were suffering from epilepsy and diabetes. While most events in the progression of anomaly were in agreement with the major trend, several exceptional events appeared in the second and the third time slots of the selected anomaly (Fig.~\ref{fig:interface}(g)).  
By splitting the sequence of the anomalous patient from others, the comparison glyphs uncovered suspicious events at these time slot. The experts noticed an abnormal lab event with a 0.59 anomaly score and 100\% support rate, CK-MB (Fig.~\ref{fig:mimic}(i)). \q{This is a critical indicator for myocardial infarction}, E1 said. The experts also found an event that was continuously missing in several time slots throughout the entire progression, Hydralazine (Fig.~\ref{fig:mimic}(b)). \q{This drug is mainly applied to patients with chronic heart failure} explained E2. \q{This may imply different causes of epilepsy. Both heart diseases can potentially cause epileptic seizure.} The experts then explored the redundant medicines highlighted in the comparison glyphs to investigate the differences in the treatment plan and surprisingly found no medicines aimed directly at curing myocardial infarction. \q{This is unusual,} E1 said, \q{It seems the patient was treated as a regular epileptic patient.} They also found a type of medicine, Pheonobarbital(Fig.~\ref{fig:mimic}(ii)) used only by the anomalous patient. \q{I believe Pheonobarbital is mainly used for neonatal and childhood seizures according to guidelines,} said E3. \q{It is rare to see this drug prescribed for a 69-year-old man.} E4 found this finding especially useful, as he commented: \q{It is a potential drug of abuse. Long-time usage can result in physical dependence, thus should be strictly controlled. I feel this system has great potential to be applied to monitor drug misuse.}

\vspace{-0.5em}
\section{Conclusion}
We have presented \name, a visual analysis technique designed to support visual anomaly detection in event sequence data. \name incorporates an unsupervised VAE-based anomaly detection model to identify anomalous sequences and events in an interpretable manner, and a visualization system with multiple coordinated views and rich interactions is provided to facilitate interpretation via one-to-many sequence comparison. We evaluate the effectiveness and usefulness of \name through a quantitative comparison of the performance of our proposed algorithm and a case study. The study results illustrate the strengths of \name and shed light on several directions for future work, including enabling integrating an associative measurement for event abnormality that considers both anomaly score and support rate, and supporting the analysis of multiple anomalous sequences simultaneously.

\vspace{-0.1cm}
\section*{Acknowledgment}
The authors thank all medical experts for their participation in the case study. This research was supported by NSFC Grant 61602306, Fundamental Research Funds for the Central Universities, the National Grants for the Thousand Young Talents in China, the NSFC Grant 61672231 and NSFC-Zhejiang Joint Fund under Grant U1609220. Nan Cao is the corresponding author. 

\bibliographystyle{IEEEtran}
\bibliography{main}

\end{document}